\documentclass[journal=jpccck,manuscript=article]{achemso}

\usepackage[version=3]{mhchem} % Formula subscripts using \ce{}
\usepackage[utf8]{inputenc}
\usepackage{array}
\usepackage{wrapfig}
\usepackage{multirow}
\usepackage{tabularx}
\author{Raphael M. Tromer}
\affiliation[State University of Campinas]
{Applied Physics Department, State University of Campinas, Campinas, SP, 13083-970, Brazil}
\alsoaffiliation[State University of Campinas]
{Center for Computational Engineering and Sciences, State University of Campinas, Campinas, SP, 13083-970, Brazil}
\author{Levi Felix}
\affiliation[State University of Campinas]
{Applied Physics Department, State University of Campinas, Campinas, SP, 13083-970, Brazil}
\alsoaffiliation[State University of Campinas]
{Center for Computational Engineering and Sciences, State University of Campinas, Campinas, SP, 13083-970, Brazil}
\author{Cristiano F. Woellner}
\affiliation[Federal University of Parana]
{Physics Department, Federal University of Parana, UFPR, Curitiba, PR, 81531-980, Brazil}
\author{Douglas S. Galvao}
\email{galvao@ifi.unicamp.br}
\affiliation[State University of Campinas]
{Applied Physics Department, State University of Campinas, Campinas, SP, 13083-970, Brazil}
\title{The Effect of Element Composition on the Structural and Electronic Properties of Carbon, Silicon, Silicon Carbide, and Boron Nitride Gyroid Schwarzites}

\usepackage{xcolor}
\begin{document}

\section{Abstract}
Schwarzites are porous structures that present negative Gaussian curvatures. Although initially proposed for carbon, in principle, schwarzites of other elements are possible. In this work, we have carried out a detailed investigation of the effect of element composition  (C, Ge, Si, SiC, and BN) on the structural, electronic and optical properties of a gyroid schwarzite structure, the so-called G688. The DFT simulations were carried out using the well-known SIESTA code. Our results showed that formation energy values are in similar range of other related allotrope structures and are thermally stable (up to 1000~K). From the electronic analyses, our results indicate that all structures, except the carbon one, present semiconductor characteristics. From the optical properties, except for the infrared region where only silicon and germanium show some absorption, the other structures exhibit optical activity only in the visible and ultra-violet regions, and all of them have large refractive index values. For reflectivity, except for Si and Ge schwarzites that reflect almost 40\% of light, the other schwarzites exhibit low reflectivity. These characteristics make them good candidates for optoelectronic applications, such as infrared/ultraviolet absorbers, and ultraviolet blockers.



\section{Introduction}

Using triply-periodic minimal surfaces as a template for atomic arrangements of crystalline materials, Mackay and Terrones~\cite{mackay_1991} proposed curved carbon nanostructures with negative Gaussian curvature, which they called schwarzites. The stability of these porous graphitic networks have been extensively investigated in the literature with respect to their structural stability~\cite{terrones_1992,lenosky_1992,barborini_2002,benedek_2003,valencia_2003,tagami_2014,owens_2016}. Their electronic 
 structure have also been investigated~\cite{phillips_1992,huang_1993,gaito_1998,park_2003,park_2010,g6888_carbon}, and even the presence of three-dimensional Dirac fermions were reported~\cite{lherbier_2014,ahn_2018}. The presence of negative Gaussian curvature has been found to drastically reduce thermal conductivity compared to other curved carbon nanostructures~\cite{pereira_2013,zhang_2017,zhang_2018}. The porous character was also exploited in applications of gas storage\cite{damascenoborges_2018,collins_2019}. But the most prominent properties of schwarzites are the mechanical ones~\cite{miller_2016, jung_2018,felix_2019}. Since their experimental synthesis is still a challenge~\cite{nishihara_2009,nishihara_2018,braun_2018,boonyoung_2019}, recent works have applied 3D-printing techniques to reproduce the schwarzites geometries, which generated structures with excellent resistance against compressive loading and ballistic impacts~\cite{sajadi_2018,gaal_2021}.

As it is well-known for the cases of diamond and graphene crystalline structures, a version of schwarzites made of other chemical elements has already been proposed in several works where the carbon atoms are replaced by boron and nitrogen~\cite{gao_2017}, silicon~\cite{laviolette_2000,zhang_2010}, and, most recently, germanium~\cite{germanium}. The stronger sp$^3$ character of silicon (Si) and germanium (Ge) bonding produces significant structural differences in all carbon nanostructured analogs. For instance, silicene and germanene membranes differ from purely flat graphene by an out-of-plane buckling of the atomic structure~\cite{ni_2012,ohare_2012,vogt_2012,dvila_2014}. This 
occurs due to the presence of the pseudo-Jahn-Teller effect~\cite{hobey_1965,bersuker_2013,perim_2014}.

Based on these ideas, we carried out a comparative study of the effect of atom composition (C, B, N, and Si) on structural stability, electronic, and optical properties of the smallest gyroid Schwarzite structure (G688)~\cite{valencia_2003,miller_2016,g6888_carbon}, afterhere called Si-S, BN-S, and SiC-S. We have considered the cases of boron nitride, silicon, and silicon carbide structures. For completeness, we have also included literature results for carbon~\cite{g6888_carbon} and germanium~\cite{germanium} ones.

\section{Methodology}

We carried out  density functional theory (DFT) simulations using the SIESTA code~\cite{Soler_2002} to investigate the structural stability of Si-S, BN-S, and SiC-S. The simulations were performed in the generalized gradient approximation {GGA-PBE}~\cite{Perdew_1996} to represent the exchange-correlation term. We considered a linear combination of atomic orbitals with a z-basis Double Zeta Valence (DZP) set basis \cite{Soler_2002} to represent the wave functions. The interaction between valence electrons and atomic ions is represented by the norm-conserving Troullier-Martins pseudopotential. The Monkhorst-Pack scheme with a $2\times 2\times 2$ k-point mesh was chosen for sampling the reciprocal space~\cite{Monkhorst_1976}, and the Brillouin zone for all structures was sampled by k-points along a simple cubic cell, as in the references~\cite{Feng_2020,germanium}. The mesh cut-off of $200$ Ry was used.

We considered as a convergence criterion that each self-consistent calculation cycle is performed until the maximum difference between  
elements of the density matrix is smaller than $10^{-4}$ eV.  The atoms and lattice vectors are optimized 
simultaneously and the force convergence criterion (in each atom) for values smaller than $0.01$ eV/\AA. 

After setting the convergence parameters through structural analysis tests, we computed the formation energy for each schwarzite structure composed of BN, Si, and SiC atoms, and for comparison purposes, we also considered
FCC BN and SiC (diamond structure), and 2D hexagonal (graphene) structures. The formation energy by an atom is given by:
\begin{equation}
E_{\text{formation}}=\frac{1}{N_{\text{total}}}\bigg (E_{\text{structure}}-\displaystyle\sum_{i=1}^{\text{Ntype}}N_iE_i\bigg ),
\label{eq:formation}
\end{equation}
where $E_{structure}$ is the total energy of the structure, $N_{type}$ is the different atoms types, $N_i$ is the number of atomic species $i$, and $E_i$ is the total energy of the isolated $i$ type atom.

In order to investigate the structural stability at high temperatures  ($T=1000$ K), we also carried out ab initio
molecular dynamics (AIMD) simulations, as implemented in the SIESTA code. We used a time step of $0.1$ fs in an NVT ensemble corresponding to a total simulation time of $2$ ps. We used 
the Nosé-Hoover thermostat to control the temperature set in the simulations.

Once the structures were fully optimized, we then carried out optical simulations with the tools available in the SIESTA code and 
using the same parameters discussed above. For each crystal direction, we assumed an external electrical 
field of magnitude $1.0$ V/\AA~, which is a typical value used in this type of simulation \cite{Fadaie2016,germanium,tromer_2020}.

The optical properties can be obtained from the complex dielectric function $\epsilon =\epsilon_1+i\epsilon_2$. The quantities $\epsilon_1$ and $\epsilon_2$ are the real and imaginary parts of the dielectric function, respectively.

From the Kramers-Kronig transformation, it is possible to derive $\epsilon_1$ as

\begin{equation}
\epsilon_1(\omega)=1+\frac{1}{\pi}P\displaystyle\int_{0}^{\infty}d\omega'\frac{\omega'\epsilon_2(\omega')}{\omega'^2-\omega^2},
\end{equation}
where $\omega$ is the frequency of the photon.

The imaginary part, $\epsilon_2$, is extracted directly from Fermi's golden rule:

\begin{equation}
\epsilon_2(\omega)=\frac{4\pi^2}{\Omega\omega^2}\displaystyle\sum_{i\in \mathrm{VB},j\in \mathrm{CB}}\displaystyle\sum_{k}W_k|\rho_{ij}|^2\delta	(\epsilon_{kj}-\epsilon_{ki}-\omega),
\end{equation}
where $\rho_{ij}$ is the dipole transition matrix element, VB and CB refer to the valence and
conduction bands, respectively, and $\Omega$ is the unit cell volume.

As discussed before, the optical properties as the absorption
coefficient $\alpha$, reflectivity $R$, and refractive index $\eta$, can be obtained directly from the real and imaginary part of the dielectric function $\epsilon_1$ and $\epsilon_2$, by the expressions:
\begin{equation}
\alpha (\omega)=\sqrt{2}\omega\bigg[(\epsilon_1^2(\omega)+\epsilon_2^2(\omega))^{1/2}-\epsilon_1(\omega)\bigg ]^{1/2},
\end{equation}
\begin{equation}R(\omega)=\bigg [\\frac{(\epsilon_1(\omega)+\epsilon_2(\omega))^{1/2}-1}{(\epsilon_1(\omega)+\epsilon_2(\omega))^{1/2}+1}\bigg ]^2 ,\end{equation}
\begin{equation}
\eta(\omega)= \frac{1}{\sqrt{2}} \bigg [(\epsilon_1^2(\omega)+\epsilon_2^2(\omega))^{1/2}+\epsilon_1(\omega)\bigg ]^{2}.
\end{equation}

\section{Results}

\subsection{Structural Parameters and Stability}
\begin{figure}
 \begin{center}
\includegraphics[width=0.8\linewidth]{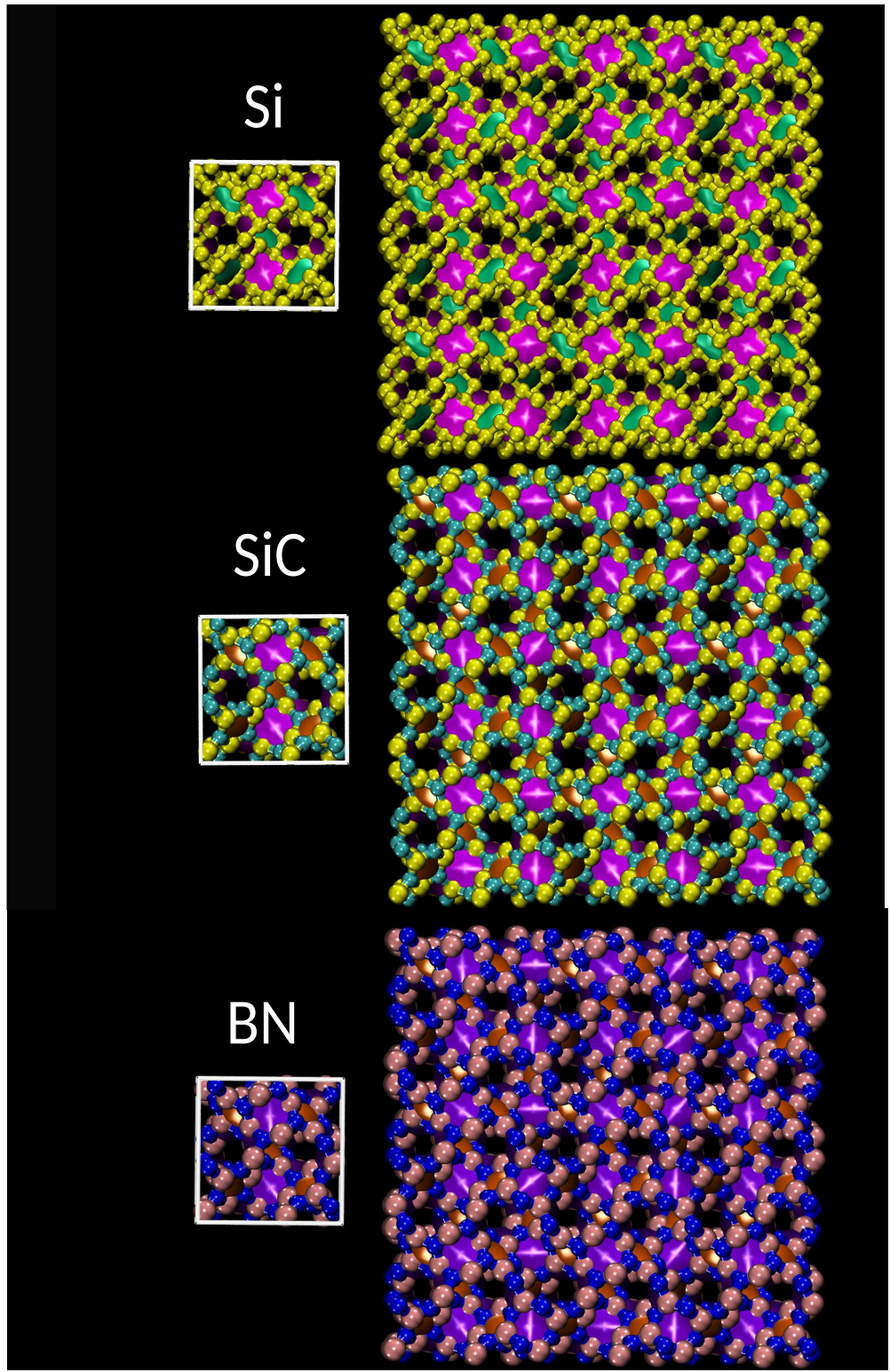}
\caption{Right: Optimized schwarzites structures. From top to bottom: Si-S, SiC-S, and BN-S. Left: The corresponding supercell units. For a better visualization of the hexagons and octagons rings, they are indicated in different colors.}
\label{fig:structure}

\end{center}
\end{figure}

In Figure \ref{fig:structure} we present the optimized Si-S, SiC-S, and BN-S structures investigated in this work. On left, we have their supercell composed of 96 atoms and on right their replication of $3\times 3\times 3$ along the x, y, and z-directions.  The G688-schwarzites consist of rings composed of six and eight atoms, represented by pink and green/red hatched regions, respectively. As can be seen from Figure \ref{fig:structure},  the structure are porous and contain large void spaces within their volume.

In Table \ref{tab:structural} we present the optimized structural parameters for Si-S, SiC-S, and BN-S structures investigated in this work. The comparison purposes, the literature corresponding results for carbon~\cite{g6888_carbon} and germanium~\cite{germanium} structures were also included. 

 All optimized structures present a square lattice symmetry, where the crystal lattice angle values are equal to $90.0^o$. The lattice vectors are: Ge (15.28 \AA), Si (14.64 \AA), and SiC (12.03 \AA), much larger than the corresponding values for the carbon ($9.66$ \AA~) and BN($9.77$ \AA~), respectively. The same trend is observed for $R_{med}$, which is the average distance between atomic bonds. The values are $2.47$, $2.32$ and $1.81$, $1.45$, and $1.47$~\AA, for Ge, Si, SiC, C, and BN, respectively.
This is a natural consequence of their van der Walls radius values, where the higher values for Si ($2.10$ \AA) and Ge ($2.11$) produce longer bond distance values than C ($1.7$~\AA), B ($1.80$~\AA), and N ($1.6$~\AA). Consequently, the volume values for silicon and germanium G688-schwarzites are larger, by almost three times, as observed in Table \ref{tab:structural}. Although silicon and germanium G688-schwarzites have similar volumes, the germanium density is much higher (more than twice), reflecting that germanium mass is almost three times higher than silicon. This is one indication that Si G688-schwarzites contain large porous regions, as a consequence of having large volume and low density.
\begin{table}
\begin{center}
\begin{tabular}{|c|c|c|c|c|c|c|c|c|c|}
\hline
Structure & a(\AA)&b(\AA)&c(\AA)&$\alpha$ ($^o$)&$\beta$ ($^o$)&$\gamma$ ($^o$)&$R_{med}$ (\AA)&V (\AA$^3$)&$\rho$ (g/cm$^3$)\\
\hline
Si&14.64&14.64&14.64&90.0&90.0&90.0&2.32&3135.16&1.43\\
\hline
SiC&12.03&12.03&12.03&90.0&90.0&90.0&1.81&1742.34&1.83\\
\hline
BN&9.77&9.77&9.77&90.0&90.0&90.0&1.47&931.85&2.12\\
\hline
C&9.66&9.66&9.66&90.0&90.0&90.0&1.45&897.05&2.13\\
\hline
Ge&15.28&15.28&15.28&90.0&90.0&90.0&2.47&3564.50&3.25\\
\hline
\end{tabular}
\caption{\label{tab:structural}Structural parameters calculated in this work for Si-S, BN-S, and SiC-S. The corresponding values for carbon and Germanium were taken from refs. ~\cite{g6888_carbon} and ~\cite{germanium}, respectively.}
\end{center}
\end{table}

In Table \ref{tab:energy} we present the formation energies values obtained from equation \ref{eq:formation} for the G688-schwarzite structures and, for comparison purposes, for other related structures.   

\begin{table}
\begin{center}
\begin{tabular}{|c|c|}    
\hline
Structure&E$_{\text{formation}}$(eV/atom)\\
\hline
Si&-5.05\\
\hline
SiC&-7.11\\
\hline
BN&-8.87\\
\hline
C&-8.80\\
\hline
Ge&-4.39\\
\hline
Silicene&-5.10\\
\hline
Si (FCC)&-5.53\\
\hline
2D SiC &-7.26\\
\hline
3D-FCC-SiC&-7.59\\
\hline
2D BN&-9.17\\
\hline
3D-FCC-BN&-8.99\\
\hline
graphene&-8.81\\
\hline
graphite&-9.1\\
\hline
Germnene&-4.40\\
\hline
Ge (FCC)&-4.63\\
\hline
\end{tabular}    
\caption{\label{tab:energy} Formation energy values of the structures investigated in this work, and for other related structures.}
\end{center}
\end{table} 

The minimum difference between the formation energy of the G688-schwarzites investigated here and their similar structures are only 0.05, 0.15, 0.12, 0.07, 0.01 eV/atom for Si-S, SiC-S, BN-S, C-S, and Ge-S, respectively. Therefore, we verified that in a possible synthesis experimental process, the G688-schwarzites investigated here present formation energy values similar to those structures that have already been synthesized. We can consider their synthesis feasibility for a possible synthetic route to G688-schwarzites could be the same proposed to carbon-based ones considering zeolites as a template \cite{nishihara_2009,nishihara_2018,braun_2018}.

Another important aspect is related to the structural changes at high temperatures. Besides verifying the structural stability, it is important to stress that there are some synthesis methods that are applicable only whether the system is stable at high temperatures, for example the fullerenes \cite{fullerene}

In order to determine whether the G688-schwarzites investigated in this work are stable at high temperatures, we carried out AIMD with an NVT ensemble during $2000$~fs at T=$1000$ K.

 \begin{figure}
 \begin{center}
\includegraphics[width=0.91\linewidth]{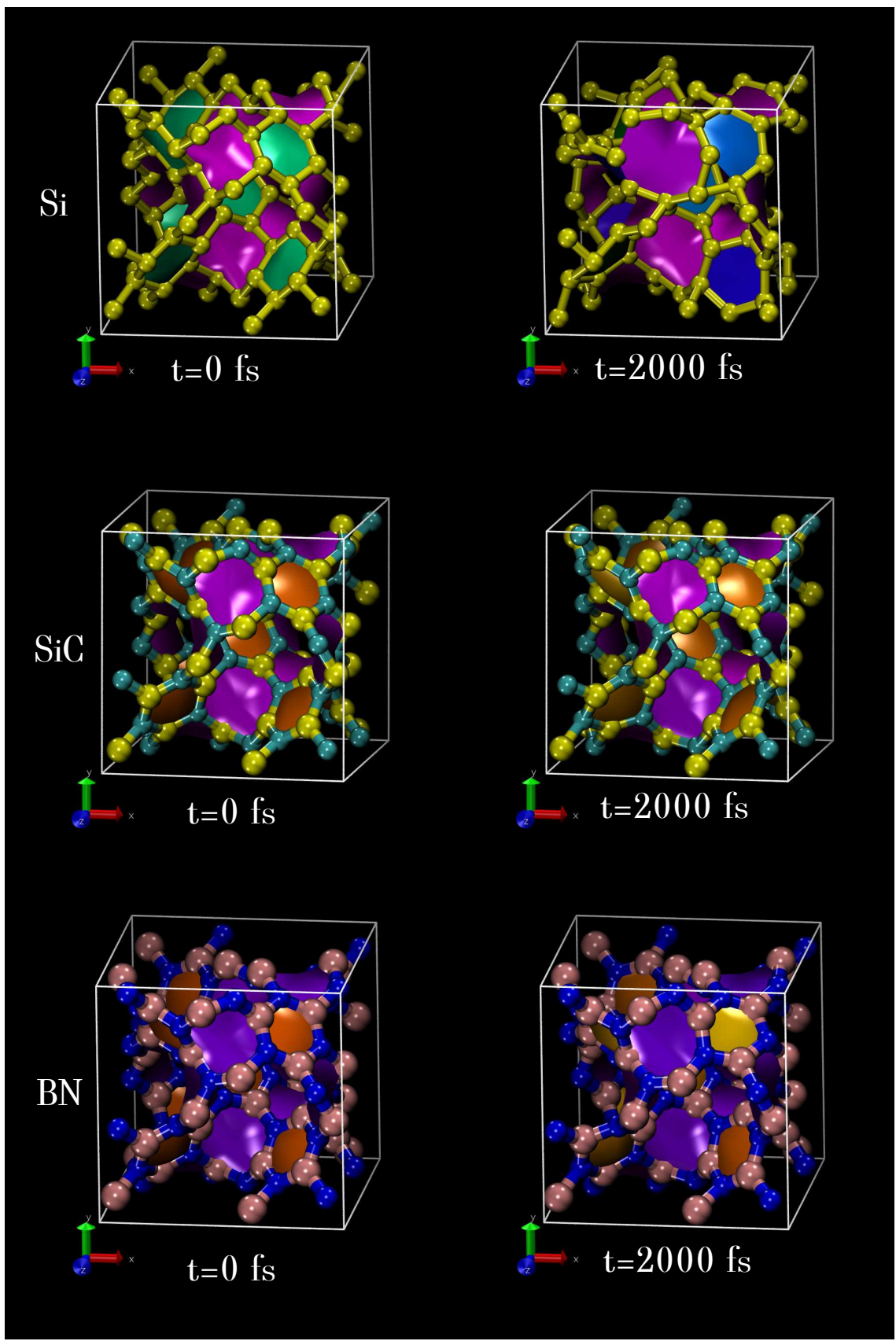}
\caption{Representative AIMD snapshots of beginning and ending of the simulations for the Si-S, SiC-S, and BN-S schwarzites.}
\label{fig:md}               
\end{center}
\end{figure} 
 
 In Figure \ref{fig:md} we present representative AIMD snapshots of the beginning and ending of simulations ( AIMD with an NVT ensemble during $2000$~fs at $T=1000$~K) for the Si-S, SiC-S, and BN-S. We can see from this Figure that all structures at $t=2000$~fs do not exhibit significant structural changes in comparison to the initial structures at $T=0$~K. There are only small structural differences between the cases, which can be attributed to thermal fluctuations. The same kind for analysis was carried out for other C and Ge G688-schwarzites and the results are similar, further evidencing the strong structural resilience of these schwarzites even at high temperatures. As mentioned above, this can be important for high-temperature synthesis methods, as zeolite-template ones \cite{braun_2018}.

\subsection{Electronic Properties}

\begin{figure}
 \begin{center}
\includegraphics[width=0.91\linewidth]{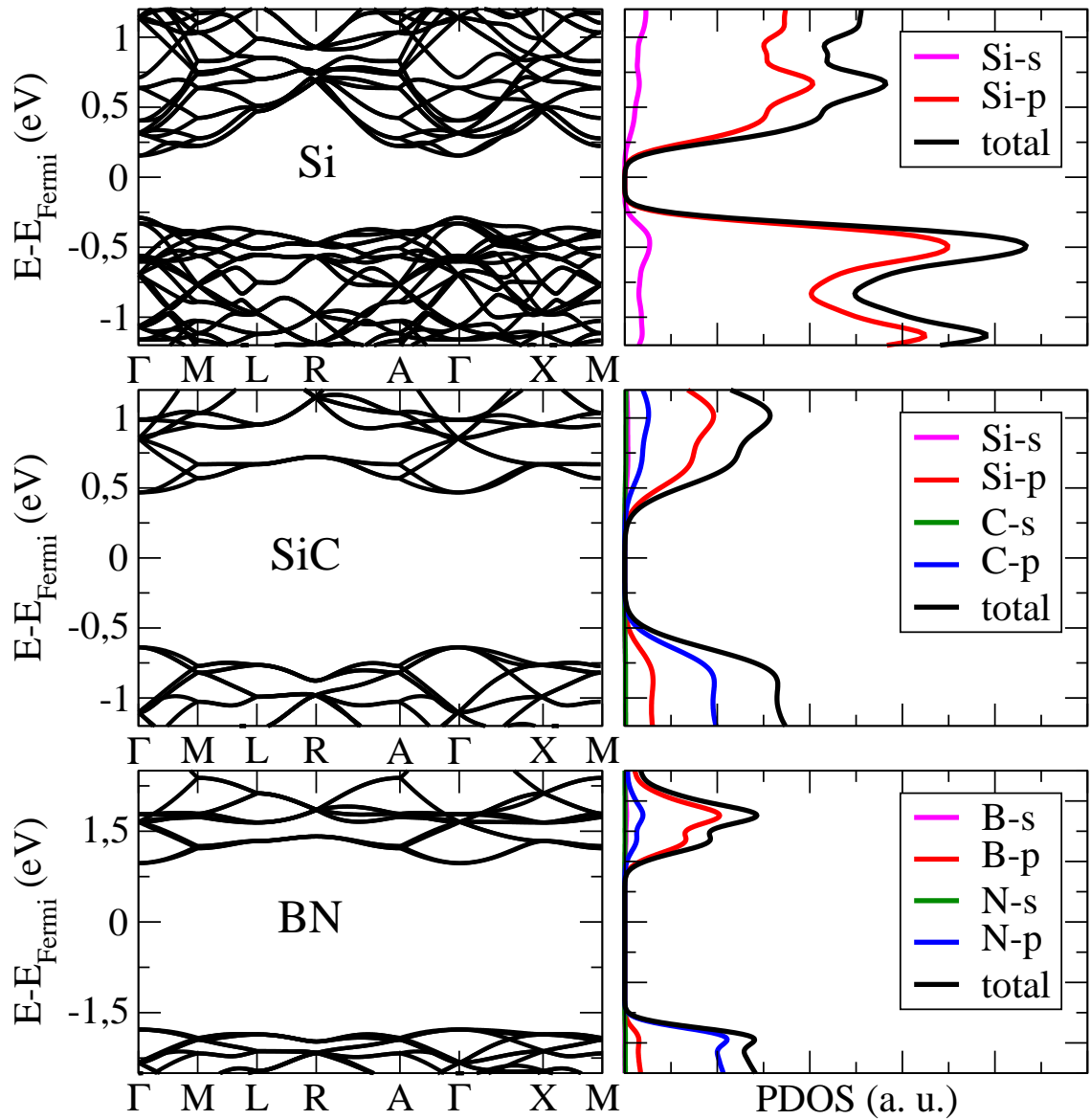}
\caption{Electronic band structure calculation (at left) and the corresponding projected density of states (PDOS) at right for the Si-S, SiC-S, and BN-S schwarzites.}
\label{fig:bands}                                               
\end{center}
\end{figure}

In Figure \ref{fig:bands}, we present the electronic band structure calculations (left) and, corresponding projected density of states (PDOS) (right) for the Si-S, SiC-S, and BN-S. We considered the same path of the irreducible Brillouin zone used in reference \cite{germanium}. For all cases, we can see that structures exhibit semiconductor characteristics with a direct bandgap from the $\Gamma$ to $\Gamma$. The values are $0.44$, $1.10$, and $2.74$ eV for Si, SiC, and BN G688-schwarzites, respectively. These values should be slightly large because it is well-known that GGA-PBE tends to underestimate the bandgap values. If we estimate that the correction is around $1$~ eV, we can conclude that Si and SiC are semiconductors and BN is a wide bandgap material. For corresponding 3D structures in the literature, we have the following values for the bandgap values: $1.1$, $3.3$, and $5.5$ eV, for Si, SiC and BN, respectively \cite{bngap,Sigap,sigap2}. The ordering of gap values of the 3D system matches the ordering of the schwarzite analogs. In Figure \ref{fig:bandsC} in the Supplementary Materials, we present the electronic band structure of the carbon-S, which is similar to the one presented in reference \cite{g6888_carbon}. This is the only case in which the system is metallic, while the Ge case is a semiconductor with a bandgap value of $0.27$~eV, investigated in the reference \cite{germanium}. The analyses of the PDOS \ref{fig:bands} show, as expected, that the p-orbitals dominate the contributions to the valence and conduction bands. For the binary system, SiC and BN, carbon and nitrogen contribute more to the valence bands, while silicon and boron to the conduction ones.

In order to illustrate the spatial distribution of the crystalline orbitals, we displayed in Figure \ref{fig:carga} of the Supplementary Material, the highest occupied crystalline orbitals (HOCO) and the lowest unoccupied crystalline orbital (LUCO).

\subsection{Optical Properties}

\begin{figure}
 \begin{center}
\includegraphics[width=0.91\linewidth]{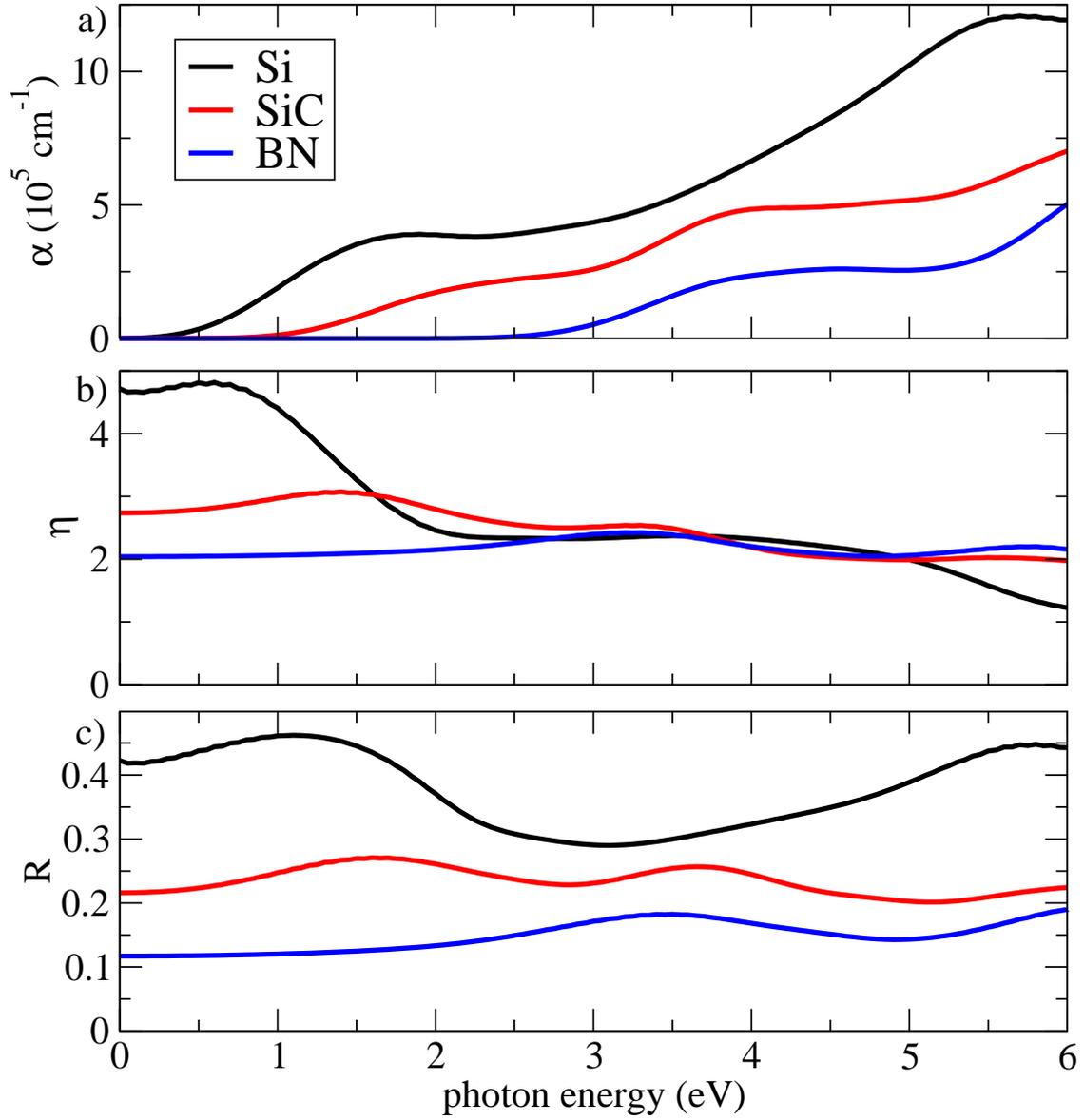}
\caption{Optical properties for Si-S, SiC-S, and BN-S schwarzites as a function of photon energy.}
\label{fig:optical}                                               
\end{center}
\end{figure}

 In Figure \ref{fig:optical} we present the optical properties for Si-S, SiC-S, and BN-S, as a function of photon energy. The corresponding results for the C-S is presented in the Supplementary Materials. The threshold absorption values ($\alpha$) are different for Si-S ($0.4$), SiC-S ($1.0$),  and BN-S ($2.5$). These values are related to each electronic bandgap value due to the first optical transition ($\Gamma$ to $\Gamma$, HOCO to LUCO) shown in Figure \ref{fig:bands}. We observed that the general trend for the G688-schwarzites is that the optical absorption starts in the infrared region (Si-S,Ge-S~\cite{germanium}, SiC-S, and C-S (see Figure S2 in Supplementary Materials). BN-S is the only exception, starting within the visible region. However, the first peak is located at the beginning or middle of the visible range for Si-S and SiC-S, respectively, while all other peaks are located in the ultraviolet region. 
 
The refractive index $\eta$ shown in Figure \ref{fig:optical}-b) show a very similar constant behavior for SiC-S and BN-S, with values close to 2, while Si-S is almost constant in the infrared region (close to 5) and decreases to 2 at the beginning of the visible range being practically the same for the other cases in the visible and ultraviolet regions. This behavior is similar to the Ge G688-schwarzite investigated in reference~\cite{germanium}. It is interesting to notice that the refractive index is independent of the chemical elements within the visible region. 

In Figure \ref{fig:optical}-c) we present reflectivity as a function of photon energy. We do not observe large variations for all cases, the behavior is practically constant within the investigated regions. The maximum variation for all cases was 0.1. The maximum reflectivity is for Si-S and the minimum for BN-S. As the reflectivity is decreased for SiC-S and BN-S in all spectrum, and small for Si-S within visible range, we can conclude that schwarzites can be used for solar cell application because the refractive index has values close to 2 within the visible range. Besides, for SiC-S and BN-S case,s the major part of light will be absorbed in the infrared and ultraviolet regions, thus these two types can be used for applications of infrared and ultraviolet absorbers. In contrast, for Si-S almost $40\%$ of light incident in the material will be reflected at infrared and ultraviolet ranges. Thus the Si case can be used for applications in ultraviolet blockers. We observed that the G688-schwarzites investigated in this work present a multi-range optical activity from infrared to ultraviolet spectrum, suggesting they could be exploited in optoelectronics applications.

\section{Summary and Conclusions}

In summary, we have carried out a detailed investigation of the effect of element composition on the structural, electronic and optical properties of a gyroid schwarzite structure, called G688. In our investigation, we considered structures composed of Ge, Si, C, SiC, and BN elements. We obtained the following values for the lattice vectors 15.28, 14.64, 12.03, 9.66, and 9.77 \AA, for Ge, Si, C, SiC and BN, respectively. This ordering is a consequence of their van der Waals radii values, where the germanium and silicon atomic volumes are larger, by almost three times, as observed in Table 1. We contrasted the formation energy of the structures investigated here with related allotropes and the obtained values are in similar range.  We also investigated the thermal stability of the different schwarzites, the results can stand high temperatures (up to 1000~K) without significant structural deformations. From the electronic analyses, our results indicate that all structures, except the carbon one, present semiconductor characteristics. From the optical properties, except for the infrared region where only silicon and germanium show some absorption, the other structures exhibit optical activity only in the visible and ultra-violet regions, and all of them have large refractive index values. For reflectivity, except for Si and Ge schwarzites that reflect almost 40\% of light, the other schwarzites exhibit low reflectivity. These characteristics make them good candidates for optoelectronic applications, such as infrared/ultraviolet absorbers, and ultraviolet blockers.

\section{Acknowledgements}

This work was financed in part by the Coordenação de Aperfeiçoamento de Pessoal de Nível
Superior - Brasil (CAPES) - Finance Code 001, CNPq, and FAPESP. The authors thank
the Center for Computational Engineering \& Sciences (CCES) at Unicamp for financial
support through the FAPESP/CEPID Grant 2013/08293-7. 

---------------------------------------------------------------------------------------------------------
%\bibliography{achemso-demo}

\providecommand{\latin}[1]{#1}
\makeatletter
\providecommand{\doi}
  {\begingroup\let\do\@makeother\dospecials
  \catcode`\{=1 \catcode`\}=2 \doi@aux}
\providecommand{\doi@aux}[1]{\endgroup\texttt{#1}}
\makeatother
\providecommand*\mcitethebibliography{\thebibliography}
\csname @ifundefined\endcsname{endmcitethebibliography}
  {\let\endmcitethebibliography\endthebibliography}{}

%\newpage
%\section{Supplementary Materials}

%\beginsupplement

% \begin{figure}
% \begin{center}
%\includegraphics[width=0.91\linewidth]{Figures/bandsC.eps}
%\caption{Electronic band structure of carbon G668-schwarzite.}
%\label{fig:bandsC}                                               
%\end{center}
%\end{figure} 

% \begin{figure}
% \begin{center}
%\includegraphics[width=0.91\linewidth]{Figures/carga.eps}
%\caption{HOCO and LUCO for Si, SiC-S, and BN-S schwarzites.}
%\label{fig:carga}                                               
%\end{center}
%\end{figure} 

% \begin{figure}
% \begin{center}
%\includegraphics[width=0.91\linewidth]{Figures/opticalC.eps}
%\caption{Optical coefficients for carbon G668-schwarzite.}
%\label{fig:opticaC}                                               
%\end{center}
%\end{figure} 
 
%This is where your bibliography is generated. Make sure that your .bib file is actually called library.bib

%%%%%%%%%%%%%%%%%%%%%%%%%%%%%%%%%%%%%%%%%%%%%%%%%%%%%%%%%%%%%%%%%%%%
%% The appropiate \bibliography command should be plaeced here.%% Notice \tht the class file automatically sets \bibliographystyle
\end{document}